\documentclass[journal]{IEEEtran}

\usepackage{amsmath}
\usepackage{amsfonts}
\usepackage{graphicx}
\usepackage{array}
\usepackage{verbatim}
\usepackage{cite}
\usepackage{url}
\usepackage{varioref}
\usepackage{hyperref}
\usepackage{cleveref}
\usepackage{color}
\usepackage{wrapfig}
\usepackage{lipsum}
\usepackage{scrextend}

\title{\LARGE \bf
Towards blockchain-based robonomics: \\autonomous agents behavior validation
}

\author{Konstantin Danilov$^{1}$, Ruslan Rezin$^{1}$, Alexander Kolotov$^{1}$ and Ilya Afanasyev$^{1}$%
\thanks{$^{1}$K. Danilov, R. Rezin, A. Kolotov and I. Afanasyev are with the Robotics Institute, Innopolis University, Innopolis, Russia, e-mail: \{k.danilov, r.rezin, a.kolotov, i.afanasyev\}@innopolis.ru}
}

\begin{document}

\maketitle
\newpage

\begin{abstract}
The decentralized trading market approach, where both autonomous agents and people can consume and produce services expanding own opportunities to reach goals, looks very promising as a part of the Fourth Industrial revolution. The key component of the approach is a blockchain platform that allows an interaction between agents via liability smart contracts. Reliability of a service provider is usually determined by a reputation model. However, this solution only warns future customers about an extent of trust to the service provider in case it could not execute any previous liabilities correctly. From the other hand a blockchain consensus protocol can additionally include a validation procedure that detects incorrect liability executions in order to suspend payment transactions to questionable service providers. The paper presents the validation methodology of a liability execution for agent-based service providers in a decentralized trading market, using the Model Checking method based on the mathematical model of finite state automata and Temporal Logic properties of interest. To demonstrate this concept, we implemented the methodology in the Duckietown application, moving an autonomous mobile robot to achieve a mission goal with the following behavior validation at the end of a completed scenario. 


\end{abstract}
\section{Introduction}
During the last decade, autonomous agents have become more intelligent and efficient in terms of the speed and the accuracy, demonstrating an impressive performance of various specific tasks and successful collaboration with each other, reaching own independent goals. This brings up the concept of the decentralized trading market where autonomous agents can consume and produce services with respect to a common protocol. Since appearance of the blockchain technology, in particular implementation of Bitcoin \cite{bitcoin}, which solved the double-spending attack problem, the decentralized multi-agent system concept gets started to be viable as far as the agents do not rely on a single point of failure and their operations are transparent for monitoring. Although the blockchain was originally introduced as the cryptocurrency solution, its idea inspired the developers of the Ethereum platform \cite{ETH17} to suggest a new concept: smart contract - an algorithmic enforcement of agreements. Very soon it became the biggest platform for decentralized applications that gave new pulse to development of such market services as selling computer's processing power and decentralized computation storages \cite{SONM2017}. Another example of a blockchain-based platform is the decentralized network of autonomous agents IOTA \cite{iota_white_paper} that is oriented to the IoT (Internet of Things) market. In spite of the fact that discussed projects become popular, they are concentrated on specific types of autonomous agents severely limiting available services. The AIRA project firstly introduced the concept of the decentralized trading market where both autonomous agents and people can consume and provide services. The authors defined the concept as "robot economics" (robonomics) ~\cite{aira_doctrine}.
This AIRA approach suggests rejecting the principle of centralized robot control to provide decentralized communication between robots and humans, using blockchain-based smart contracts as a basis of these communications ~\cite{aira_main}. Since liabilities depend on real world processes there is no guarantee that agents will always execute them correctly. It can be caused by a service provider due to (1) an intentional fraud or (2) an incorrect behavior because of malfunctions. For the first problem AIRA developers suppose that a reputation model can be developed and integrated into the consensus protocol that will reflect in blockchain information about the extent of trust to the agent. Thus, consumers can further rely on this information for selection of a service provider. For the second problem AIRA project still requires both technical and theoretical solutions, which are the focus of the current paper.
In this work we concentrate our attention on the agents whose behavior can be described by a nondeterministic finite state automata \cite{RAB59} (further called as agent-based systems). For these agents we present the novel validation methodology of liability execution that allows detecting malfunctioned agents. It works with the assumption that a malfunctioned service provider produces a result that contradicts its behavioral model and leads to automatic suspension of purchase. The validation approach is based on the formal software verification method, Model Checking \cite{CLA99}. Since mobile robots can be used in the real life, e.g. in logistics, transportation, etc., we provide the simple prototype implementation for a Duckietown environment~\cite{duckietown}, where a mobile robot moves through the town, oriented by a given tags’ sequence, to achieve a mission goal with its behavior validation at the end of a completed scenario. Our validation approach can be implemented (1) into a consensus protocol directly, or (2) as a part of a decentralized blockchain application. In the former case it is assumed that the validation of liability execution could be done in a decentralized form on validator nodes (miners) and transactions, confirming that the service was provided in a proper manner, will be included in the new block. For instance, it could be implemented with existing solutions supporting pluggable consensus like Parity or Hyperledger~\cite{hyperledger}. In the latter case it is integrated into the AIRA's approach, where the validation is performed by a third-part application, which commits the validation result to the Ethereum smart-contract.
%
%

The paper is organized as follows: Section \ref{sec:related_work} considers related works, Section \ref{sec:methodology} formalizes our methodology, Section \ref{sec:implemented_prototype} describes the methodology implementation in a scenario with a mobile robot motion in the Duckietown environment with the blockchain-based validation of robot's behavior. Finally, we present our results in Sections \ref{sec:result}, discuss the concept in Section \ref{sec:discussion} and conclude in Section \ref{sec:conclusion}.
\section{Related work}
\label{sec:related_work}
In this section we review model formalization and verification works, that are helpful for understanding the proposed solution, which is described in \ref{sec:methodology}. 


The work \cite{WILL89} considers a formalization of the model 
as a tuple $(\mathfrak{U}, \mathfrak{B})$, where $\mathfrak{U}$ is the universum set of all possible outcomes and $\mathfrak{B}$ is a  behavioral set of allowed outcomes. The parameterization of models is studied by introducing a canonical form and invariant notions. Although the work \cite{WILL89} mostly concentrates on time invariant linear dynamical systems nevertheless applications of the approach to the discrete-event systems (DES) are also considered. Moreover, the author introduces the interconnection approach to study subsystems separately from the whole system. Finally, the work demonstrates how models can be built from the observed data through "most powerful unfalsified models" (MPUM) approach.

In the work \cite{ROL15} authors provide a review for the modeling of the multi-unmanned vehicle mission, describing the Agent-based model, Business process model notation (BPMN), Petri Net (PN), Hidden Markov Models (HMM), State Machine (SM) and Tree Model (TM). The authors concluded that PN and HMM are the most appropriate techniques for the modeling in robotics applications. Nevertheless, as we show in Section \ref{sec:methodology}, agent-based models are also interesting and promising for multi-robot systems. Another our observation is that although the paper \cite{ROL15} considered BPMN, it did not include models based on Process calculus such as Communicating Sequential Processes (CSP) \cite{HOA85}, Calculus of Communicating Systems  (CCS) \cite{MIL80} and Pi-Calculus \cite{MIL99}.

In the work \cite{LYO15} an application of the Process calculus for description and verification of a robot behavior is studied. The advantage of using Process calculus over Model Checking technique is that the "state explosion problem" can be avoided however new challenges such as checking equivalence between processes must be considered. The authors also state that in case using Dynamic Bayesian Network (DBN), the structural equivalence problem can be reduced to a filtering problem in terms of DBN.

The blockchain looks very promising as a core technology to build a service market for autonomous robots. In the paper \cite{blockchain_swarm} it is proposed to use Bitcoin to perform distributed decision making through voting mechanism, which is implemented by using agents depositing a "decision account". Ethereum project \cite{ethereum_white_paper} has introduced more reliable consensus protocol and smart contracts, allowing to implement more flexible logic of an interaction between autonomous agents than in Bitcoin due to the Turing Complete programming language. Although the scientific community actively works on Bitcoin improvements, which mostly addressed to disadvantages of the blockchain technology such as scalability \cite{sharding, plasma}. However, the decentralized trading market requires an approach that can validate a liability of the autonomous agents based on one of the representations of behavior model discussed above or new one. This can help to expand the blockchain technology from financial flows to proving events, which occur in real market like a service placement or consumption.
\section{Methodology}
\label{sec:methodology}
Consumption of services in a decentralized trading market is controlled by blockchain-based smart contracts. The technology connects a customer to a service provider without any necessity to trust each other. Moreover, communicating parts can be either humans or autonomous agents. Although smart contracts are written using the Turing complete language, computations related to the validation of the service execution are not applicable due to high complexity and cost. Therefore, a third member is introduced in the contract, which is responsible for the validation. It should be noticed that in general it is not required: for instance, if the model validation can be integrated into the consensus mechanism. 

\begin{figure}[ht!]
\includegraphics[width=\linewidth]{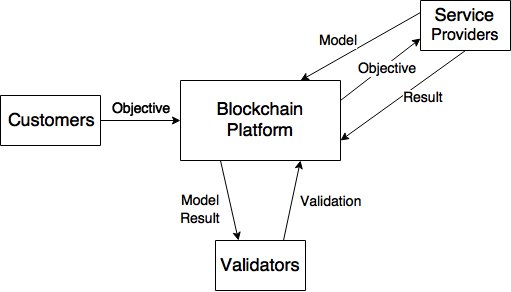}
\caption{
General scheme of interaction between autonomous agents during the liability lifecycle. A customer orders a service by submitting a transaction with an objective. A service provider takes the liability to satisfy the objective and submits the transaction with the execution result on completion. Validators check the correctness of the liability execution.
}
%
%
%
%
%
\label{fig:rliab}
\end{figure}

The high-level block-scheme of an interaction between parts via smart contract is shown in Fig. \ref{fig:rliab}. At the first step, a smart contract is created with an objective from a customer and a corresponding behavior model from a service provider. After that the service provider starts executing liability according to its behavior model. Simultaneously, this provider records a log (e.g., GPS path). In AIRA concept this log is considered as a result. When the service is finished, the result is submitted to the smart contract that confirms its work. Finally, a validator must check correctness of the liability execution, based on the behavior model and the result, and only after this step the service provider will be paid. Usually an objective is a specific service, therefore a service provider can supply a customer with a guide or an interface for objective construction. However, there is a necessity in a method that formally defines a behavior model, a liability execution result and a validation procedure. Due to the requirement that validators must automatically check any liabilities permitted in the market, the validation procedure must not depend on service.
%
%

%
%
%

In the current work we consider only such service providers that execute computations based on the finite number of states. States are defined by values of variables that are changed as a computation progresses. In general, such services consist of communicating devices that interact according to a protocol and can be studied as agent-based systems. We use the methodology of such systems defined in the work \cite{REZ18}. Although the authors applied this methodology to multiplayer computer games the idea can be generalized for an arbitrary finite state protocol. Similarly, we call agent-based system a tuple $\mathcal{G} = (\mathcal{A}, \mathcal{V}, \mathcal{V}_{init}, c, \mathcal{OP}, \mathcal{EV})$, where $\mathcal{A}$ is a set of actors; $\mathcal{V}$ is a set of vectors of attributes; $\mathcal{V}_{init}$ is a set of possible initial vectors of attributes; $c$ is a tuple of parameters; $\mathcal{OP}$ is a function that maps set of actions to each actor; and $\mathcal{EV}$ is the evolution operator that defines rules of changes of the system. The meaning of the definition is following: after fixed period of time every agent must submit an action. The actions are performed simultaneously changing the attributes that describe the state of agent-based system. The $c$ tuple is used to parametrize actions and domain of the attributes. Such parametrization can help to deal with "state explosion problem" during verification at applying formal methods such as Model Checking \cite{CLA99}. The main advantage of the methodology is a representation of agent-based system (i.e. a service provider behavior), as a program in Model Checker language. Such representation is referred as a \textit{model} hereafter. Moreover, we contribute with an extension of the methodology representation that allows providing a validation of the service liability based on the model and a result.

Model Checker is a special tool with already implemented Model Checking algorithm that takes as input a symbolic representation of finite state automata in a special language (the \textit{model}) and Temporal Logic \cite{CLA86} formula as property of interest. In result, Model Checker answers the question whether a property holds on the model, meaning that computations providing by the model are correct in some sense. If a model and a result are submitted to the smart contract by a service provider, then the task of a validator is to construct a property and to run the Model Checker. Finally, the validator either confirms or rejects liability execution based on Model Checking result. However, there are two problems that must be solved.
Firstly, Model Checker language requires a representation of automaton's states and transitions by formulas in first-order predicate logic. For big agent-based systems such procedure is not straightforward and error-prone if a service developer performs it by hand.
Secondly, it must be finalized a format of liability execution result and procedure of how the validator can generate a property of interest from it. It was decided that it will be required from agent-based system to submit a log of its execution as a result. The log must reflect changes of state of the agent-based system $\mathcal{G} = (\mathcal{A}, \mathcal{V}, \mathcal{V}_{init}, c, \mathcal{OP}, \mathcal{EV})$. It can be achieved by providing values for all variables from $\mathcal{V}$ per row. Hence each row of the file describes a state. However, a validator still needs a tool that can automatically generate a property from the log file.

We have developed a framework that includes instruments to overcome both problems, which will be discussed in details in the Section \ref{sec:implemented_prototype}.

\section{Implementation}
\label{sec:implemented_prototype}
\begin{figure}
\centering
\includegraphics[width=\linewidth]{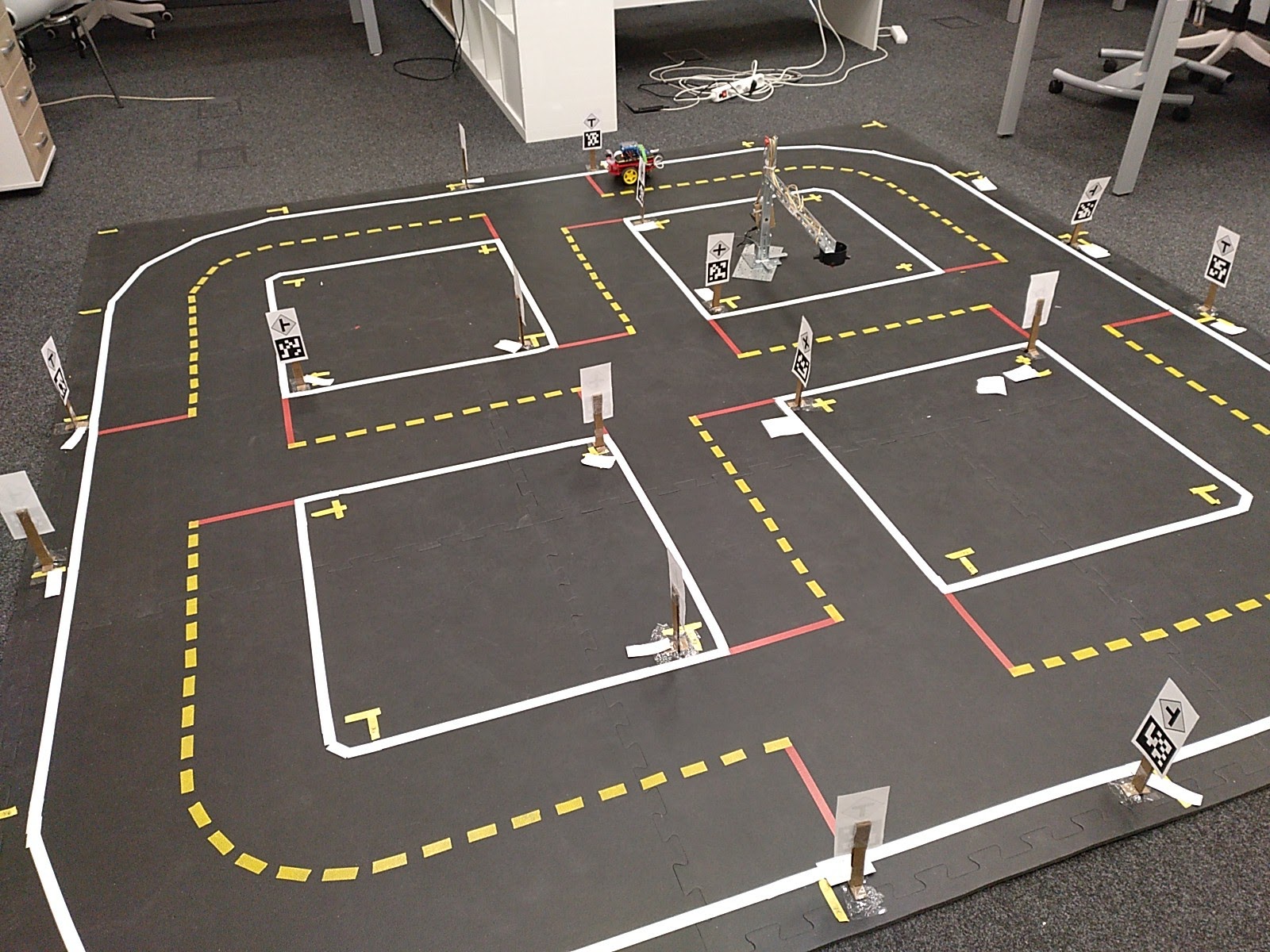}
\caption{
Our experiment in Duckietown environment. The Duckiebot executes a mission, going through "the city"  by following tags' sequence from the liability smart contract and recording observed tags into the log, which is verified after end-of-mission by a validator. Video is available at \url{https://youtu.be/mmC8jxXWWJw}
}
\label{fig:duckietown_real}
\end{figure}

To demonstrate our methodology in hardware, we implemented our concept in the Duckietown framework, considering the whole AIRA robot liability lifecycle: creation, execution and validation ~\cite{aira_main}. Thus, our prototype consists of Duckietown~\cite{duckietown}, a Duckiebot and software, which helps a user to create a scenario assigning robot’s route in the Duckietown and checking the liability execution. The liability will be executed by the Duckiebot and verified by a validator, using the approach, described in ~\ref{sec:methodology}.
We used Duckietown environment with a size of 3x3 meters (or 5x5 tiles in terms of original Duckietown project), having tiles with 4 turns, 4 T-junctions, a four-way intersection and AprilTag signs~\cite{olson2011} at each intersection that can be seen in Fig. \ref{fig:duckietown_real}.

\par

Since our research does not focus on Duckiebot navigation and localization, we used a simplified robot behavior control with tags’ sequence-based movement. Thus, the following simple algorithm for Duckiebot movement was applied:
\begin{enumerate}  
\item Set the tag sequence (e.g. 1 $\rightarrow$ left, 2 $\rightarrow$ right, 3 $\rightarrow$ forward)
\item Start the robot movement and stand before a stop line
\item Find the rightmost tag and recognize its ID
\item Find it in the sequence. If it exists, execute an assigned action and go to the step 2.
\end{enumerate}

Using this algorithm, it is possible to create a full-cycle prototype, since robot is able to perform any user sequences, recording a log and sending it as the result.

During the prototype creation, we used some ROS (Robot Operating System) nodes both original for Duckietown and our own. Developed ROS nodes realize the following behavior:

\begin{enumerate}  
\item Bot waits for new liability smart-contract by filtering events of smart contract factory with predefined address.
\item Bot gets new liability and waits an objective to be set.
\item Bot downloads the objective description (tag sequence) from IPFS and starts moving in accordance with this sequence, saving all observed tags.
\item When the sequence is finished, the bot saves a result file to IPFS and publishes its hash to the smart contract.
\end{enumerate}

Several client scripts were developed in order to provide liability creation and validation. Smart contracts were used from the AIRA core repository \cite{aira_core}. Additionally one new contract was developed  - RobotLiabilityFactory, which creates RobotLiability instances and raises the appropriate event.

The "Make order" script accepts promisor and promisee addresses (in terms of Ethereum), actions sequence and creates a liability, an objective as a JSON-file and Model Checker file, than publishes it to IPFS and its hash to the created contract. In a such way, the contract between the user and the car is created.
The "Validator" script checks for any contract, created by the RobotLiabilityFactory, waits for a result, loads it from the IPFS and passes to the Model Checker. Depending on the Model Checking result, liability is either confirmed or rejected.

During the case study the validation framework was developed in Python language. PRISM \cite{PRISMA2017} Model Checker was selected to work in conjunction with the framework and execute Model Checking algorithms. Main purpose of the tool is to provide not only validation of the liability execution but also generation of the model for Model Checking from the template and settings, construction of the Temporal logic property from the log file and construction of reduced models of the service behavior. The last feature can be useful to deal with state explosion problem.

Template file contains code in PRISM language representing the behavioral model, but explicit values of actions and parameters are substituted with tags. Tag is a valid identifier (can contain ascii letters or \_) surrounded with @ symbol (for instance, @moved@). This approach allows to avoid providing complex formulas for actions by hand and linking parameters of the model dynamically. Formulas for the tags are generated by Generator class that is obligated to be written by developer of the service in Python language. The main advantage is that traditional test driven development can be utilized to improve reliability.

Settings file contains text in YAML (YAML Ain't Markup Language) \cite{YA09} format defining arguments needed to generate values for parameters and actions from the template. Also settings file can contain a list of (key, value) pairs to provide default values for parameters and actions. Using the fact that actions are represented by first order predicate logic formulas we can provide constant default expression (true or false) for those actions which we know don't influence correctness of the property but from the other hand contribute to the number of states of the model. The strategy helps to reduce size of the model and increase verification speed.
%
%
%

\begin{figure}[ht!]
\includegraphics[width=\linewidth]{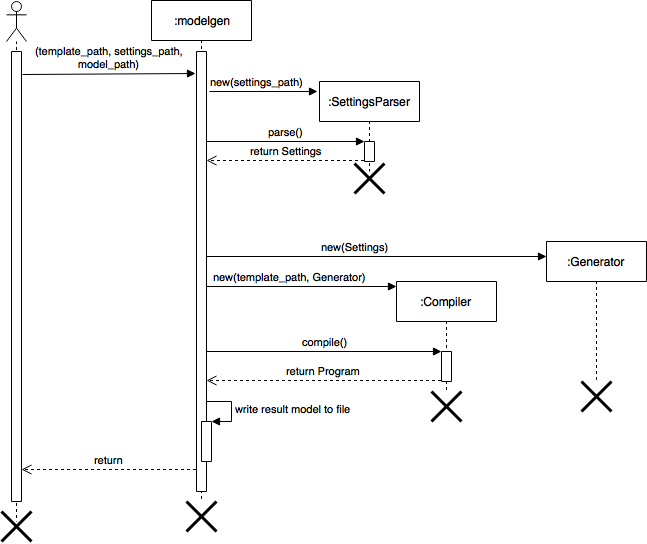}
\caption{
The dynamic view representation of a model generation tool, which compiles the final model from a template file taking into account settings.
}
\label{fig:mgen}
\end{figure}

The dynamic view of model generation tool that is a part of validation framework is shown in Fig. \ref{fig:mgen}.

To construct Temporal Logic properties from a log file another tool was developed as part of the validation framework. The tool assumes that variables of the agent-based system are enumerated and named as follows $V = \{v_1, v_2, \dots, v_m\}$. The value of $i$-th attribute at time $t$ is denoted as $v_{i}(t)$. The log file contains $n$ rows and $m$ columns. Each value is referenced as $r(i, j) \quad i = \overline{1,m},\quad j = \overline{1,n}$. $A$,$G$,$E$,$X$ are standard operators of Temporal Logic. Based on this assumption algorithm is implemented and for now allows construction of two types of properties defined recursively:
\begin{equation}
S(i) = \begin{cases}
c(n - i + 1) \wedge E(X(S(i - 1))) & 2 < i \leq n \\
c(n-1) \wedge A(G(c(n))) & i = 2\\
\end{cases}
\end{equation}
\begin{equation}
W(i) = \begin{cases}
c(n - i + 1) \wedge E(F(W(i - 1))) & 2 < i \leq n \\
c(n-1) \wedge A(G(c(n))) & i = 2\\
\end{cases}
\end{equation}

where $c(i) = \displaystyle\bigwedge_{0<j \leq m}{v_{j}(i)=r(i, j)}$. 

First property is called \textit{strong} and can be read as follows: starting from the initial state represented by the first row there is exist a transition that moves agent-based system to the state represented by the second row and so on, finally, agent-based system comes to the state represented by the last row and stay in it forever. 

The second property is called \textit{weak} and can be read as follows: starting from the initial state represented by the first row there is a sequence of intermediate transitions that moves agent-based system to the state represented by the second row and so on, finally, agent-based system comes to the state represented by the last row and stay in it forever. This property type can be useful in situations when it is not possible for agent-based system to log each change of state however the customer is satisfied enough with partial log validation.
\section{Experiment and Results}
\label{sec:result}
The experiment was performed using Duckietown infrastructure (described in the Section \ref{sec:implemented_prototype}) with a Duckiebot and a desktop computer (with Ubuntu 16.04), which were connected to each other wirelessly via the Duckietown Wi-Fi.

Regular Duckietown project nodes (e.g. line detector and line following node) use almost all CPU resources, which are caused by the fact that Duckiebot is based on Raspberry Pi3. When we try to start the IPFS Daemon in parallel with these nodes, the Duckiebot becomes unstable (lost road marking, passed the stop line, etc.). To eliminate this effect, both IPFS and Parity services were executed on the desktop computer, located in the same network. Thus, in the case of using robots with low-performing hardware, these optimization issues should be solved. Another solution is to create an infrastructure with dedicated IPFS and Ethereum services. In that case, we should emphasize security-related issues.
%
%

Although, there were some performance issues, running on the bot, several attempts were performed in the configuration with IPFS, running on the bot. During them it was observed, that there was time lag between file publication and file becomes available on other node. Worse yet, current IPFS API implementation for python has no timeout control mechanism, making work with IPFS potentially unstable. 

The robot was put on the first intersection and the system started. Messages from meaningful for this experiment nodes, as well as from demo scripts were displayed. IPFS and Parity were running on the desktop computer. Parity was started without a signer module, with 3 different unlocked account (for client, car and validator) in Ropsten network. Then the validator script was started and the ``Make order`` script was executed. The tobot got its objective and passed the city, according to the given sequence. In the end it successfully uploaded the result to the IPFS, which was checked by the validator, which in its turn issued the confirm() transaction. The experiment was recorded, the video is available on YouTube \cite{first_prototype_demo}.

Although there were some issues, the experiment should be evaluated as successful. The full liability life-cycle was shown on real hardware.
\section{Discussion}
\label{sec:discussion}
This work is a proof-of-concept of the full-cycle liability execution, including the validation stage. Having this proof-of-concept, now it is important to evaluate existing robonomics infrastructure and to propose improvements. The important part of this work is the real life case analysis. In this work we consider the simple behavior model, which can be evaluated as the first level of the complex taxi-service prototype and as the beginning point of the case study research. It is possible to sequentially increase its complexity by adding new components. Each component and case are a subject for further investigations. Open issues are the following:
\begin{enumerate}  
\item 
Usually users ask a taxi service to pick them from the point A and to the point B. It should be defined, what is the objective for the full cycle (movement to A and then to B) and how many smart-contracts should we create for this case.
\item 
In our simple example a user makes a deal directly with the robocar, that means the user should know car's location and its address in advance. However, it is not usually true. Some kind of an intermediate service is required to connect to the user and the service provider.
\item 
With the increasing number of IoT devices and road infrastructure improvement it is possible to turn them to the part of robonomic. For instance, a car can make deal with the infrastructure in order to build an optimal route or even to get active support from the infrastructure.
\item 
The AIRA liability validation model assumes that robots do not forge a log of objective execution intentionally. It means, that there is an option for service provider to get payment for not provided service. Possible solution is to introduce "recorders", for instance, which can record to the blockchain the fact of presence of an agent in the given place at the given time. This information could be considered during the log validation process. The full process description should be developed as well as economic model: someone should pay for this data.
\end{enumerate}

Obviously, complex agents behavior will require improvements in the modeling and validation framework:

\begin{enumerate}
\item The main drawback of the Model Checking approach is that the number of states of the agent-based system exponentially grows with the number of variables. This means that not all models can physically be processed and there is a need in a mechanism that restricts complexity of models.
\item Additional work must be done to expand the approach to conditions when the probabilistic model is more suitable than the nondeterministic one. In that case the Probabilistic Model Checking can be utilized.
\item  The validation framework can also be enhanced. In particular the implementation of the tool that allows generation of high level code from the Model Checking model automatically and its integration into the service development project.
\item The property generation tool can be enhanced with new property types. Moreover, usage of the tool can be extended to customers allowing them to construct their own properties, for example, in order to verify that the submitted model satisfies their needs before allowing liability to be executed by a service provider.
\end{enumerate}

\section{Conclusion}
\label{sec:conclusion}

In this work we considered one concept of a model of the decentralized trading market for autonomous agents called robonomics. For special agent-based systems the new approach based on Model Checking formal software verification technique was introduced to address validation problem of a liability execution that can help to suspend payments to malfunctioned service providers and can be integrated together with a reputation model into a blockchain consensus protocol. The validation procedure tries to prove that a result submitted by a service provider after the execution corresponds to the behavior model submitted beforehand. However, to pass a property as an input to Model Checker, it must be transformed to a Temporal Logic formula. Moreover, the construction of a model for complex systems, which is suitable for an execution by the Model Checker, is also complex task. Therefore, we have developed the framework that can help to overcome both problems. The validation approach was evaluated as a part of the complex AIRA robonomics prototype, which was implemented, using the Duckietown project. The prototype implements the simple real-life case - a robot driving according to the given route with the following liability validation. Finally, we opened the discussion about future implications for proposed solutions.
\section{Acknowledgements}
\label{sec:acknowledgements}
We would like to thank Prof. Nikolaos Mavridis for the possibility to use Duckietown environment at Innopolis University, and AIRA lab. researchers for valuable discussions and consulting in AIRA project.
\bibliography{main}

\begin{thebibliography}{10}

\bibitem{aira_core}
Aira core repository.
\newblock \url{{https://github.com/airalab/core}}.

\bibitem{aira_doctrine}
Aira: Robot economics doctrine.
\newblock
  \url{http://aira.life/wp-content/uploads/2017/08/Aira\_robot\_economics\_doctrine\_1-3.pdf}.

\bibitem{first_prototype_demo}
Duckietown aira prototype.
\newblock \url{https://youtu.be/mmC8jxXWWJw}.

\bibitem{hyperledger}
Hyperledger architecture, volume 1.
\newblock
  \url{www.hyperledger.org/wp-content/uploads/2017/08/Hyperledger\_Arch\_WG\_Paper\_1\_Consensus.pdf}.

\bibitem{YA09}
O.~Ben-Kiki, C.~Evans, and B.~Ingerson.
\newblock Yaml ain’t markup language.
\newblock \url{http://www.yaml.org/spec/1.2/spec.html}, 2009.

\bibitem{sharding}
V.~Buterin.
\newblock Ethereum sharding faq.
\newblock \url{https://github.com/ethereum/wiki/wiki/Sharding-FAQ}, 2017.

\bibitem{ethereum_white_paper}
V.~Buterin et~al.
\newblock Ethereum white paper.
\newblock \url{https://github.com/ethereum/wiki/wiki/White-Paper}, 2013.

\bibitem{CLA86}
E.~M. Clarke, E.~A. Emerson, and A.~P. Sistla.
\newblock Automatic verification of finite-state concurrent systems using
  temporal logic specifications.
\newblock {\em ACM Trans. Program. Languages \& Systems (TOPLAS)},
  8(2):244--263, 1986.

\bibitem{CLA99}
E.~M. Clarke, O.~Grumberg, and D.~Peled.
\newblock {\em Model checking}.
\newblock MIT press, 1999.

\bibitem{blockchain_swarm}
E.~C. Ferrer.
\newblock The blockchain: a new framework for robotic swarm systems.
\newblock {\em arXiv preprint arXiv:1608.00695}, 2016.

\bibitem{HOA85}
C.~Hoare.
\newblock {\em Communicating Sequential Processes}.
\newblock Prentice-Hall International Series in Computer Science. Prentice
  Hall, 1985.

\bibitem{aira_main}
A.~Kapitonov, S.~Lonshakov, A.~Krupenkin, and I.~Berman.
\newblock Blockchain-based protocol of autonomous business activity for
  multi-agent systems consisting of uavs.
\newblock In {\em Workshop on Research, Education and Development of Unmanned
  Aerial Systems (RED-UAS)}, pages 84--89, 2017.

\bibitem{PRISMA2017}
M.~Kwiatkowska, G.~Norman, and D.~Parker.
\newblock Prism 4.0: Verification of probabilistic real-time systems.
\newblock In {\em Int. conference on computer aided verification}, volume 6806,
  pages 585--591. Springer, 2011.

\bibitem{LYO15}
D.~M. Lyons, R.~C. Arkin, S.~Jiang, T.~M. Liu, and P.~Nirmal.
\newblock Performance verification for behavior-based robot missions.
\newblock {\em IEEE Transactions on Robotics}, 31(3):619--636, June 2015.

\bibitem{MIL80}
R.~Milner.
\newblock A calculus of communicating systems.
\newblock {\em Lecture Notes in Comput. Sci. 92}, 1980.

\bibitem{MIL99}
R.~Milner.
\newblock {\em {Communicating and Mobile Systems: The Pi-Calculus}}.
\newblock Cambridge University Press, Cambridge, UK, 1999.

\bibitem{bitcoin}
S.~Nakamoto.
\newblock Bitcoin: A peer-to-peer electronic cash system.
\newblock 2008.

\bibitem{olson2011}
E.~Olson.
\newblock Apriltag: A robust and flexible visual fiducial system.
\newblock In {\em IEEE International Conference on Robotics and Automation
  (ICRA)}, pages 3400--3407, 2011.

\bibitem{duckietown}
L.~Paull, J.~Tani, H.~Ahn, J.~Alonso-Mora, L.~Carlone, M.~Cap, Y.~F. Chen,
  C.~Choi, J.~Dusek, Y.~Fang, et~al.
\newblock Duckietown: an open, inexpensive and flexible platform for autonomy
  education and research.
\newblock In {\em IEEE Int. Conference on Robotics and Automation (ICRA)},
  pages 1497--1504, 2017.

\bibitem{plasma}
J.~Poon and V.~Buterin.
\newblock Plasma: Scalable autonomous smart contracts.
\newblock {\em White paper}, 2017.

\bibitem{iota_white_paper}
S.~Popov.
\newblock The tangle.
\newblock \url{https://iota.org/IOTA\_Whitepaper.pdf}, 2016.

\bibitem{RAB59}
M.~O. Rabin and D.~Scott.
\newblock Finite automata and their decision problems.
\newblock {\em IBM J. Res. Dev.}, 3(2):114--125, Apr. 1959.

\bibitem{REZ18}
R.~Rezin, I.~Afanasyev, M.~Mazzara, and V.~Rivera.
\newblock Model checking in multiplayer games development.
\newblock {\em IEEE International Conference on Advanced Information Networking
  and Applications (AINA)}, 2018.

\bibitem{ROL15}
J.~J. Rold{\'a}n, J.~del Cerro, and A.~Barrientos.
\newblock A proposal of methodology for multi-uav mission modeling.
\newblock In {\em IEEE Mediterranean Conference on Control and Automation
  (MED)}, pages 1--7, 2015.

\bibitem{SONM2017}
SONM.
\newblock Supercomputer organized by network mining.
\newblock \url{https://sonm.io}, 2017.
\newblock Whitepaper.

\bibitem{WILL89}
J.~C. Willems.
\newblock Models for dynamics.
\newblock In {\em Dynamics reported}, pages 171--269. Springer, 1989.

\bibitem{ETH17}
G.~Wood.
\newblock Ethereum yellow paper.
\newblock \url{http://yellowpaper.io}, 2014.

\end{thebibliography}
\bibliographystyle{abbrv}

\end{document}